\documentclass[12pt]{article}
\usepackage{amssymb}
\usepackage{amsmath}

\usepackage{graphicx}
\usepackage{epstopdf}
\usepackage{subfig}

\newcommand{\be}{\begin{eqnarray}}
\newcommand{\ee}{\end{eqnarray}}
\newcommand{\non}{\nonumber}

\newcommand{\qq}{\ensuremath{\mathsf{q}}}

\begin{document}

\begin{titlepage}
\strut\hfill UMTG--280
\vspace{.5in}
\begin{center}

\LARGE Boundary energy of the open XXX chain\\[0.2in]  
\LARGE with a non-diagonal boundary term\\
\vspace{1in}
\large Rafael I. Nepomechie \footnote{nepomechie@physics.miami.edu}
and Chunguang Wang \footnote{c.wang22@umiami.edu}\\[0.8in]
\large Physics Department, P.O. Box 248046, University of Miami\\[0.2in]  
\large Coral Gables, FL 33124 USA\\

\end{center}

\vspace{.5in}

\begin{abstract}
   We analyze the ground state of the open spin-1/2 isotropic quantum spin
   chain with a non-diagonal boundary term using a recently proposed
   Bethe ansatz solution.  As the coefficient of the non-diagonal
   boundary term tends to zero, the Bethe roots split evenly into two
   sets: those that remain finite, and those that become infinite.  We
   argue that the former satisfy conventional Bethe equations, while
   the latter satisfy a generalization of the Richardson-Gaudin
   equations.  We derive an expression for the leading correction to
   the boundary energy in terms of the boundary parameters.
\end{abstract}

\end{titlepage}

\setcounter{footnote}{0}

\section{Introduction}\label{sec:intro}

Ever since the open spin-1/2 XXX (isotropic) quantum spin chain with
non-diagonal boundary terms was shown to be integrable
\cite{Sklyanin:1988yz,Ghoshal:1993tm,deVega:1993xi}, the challenge has
been to find its general Bethe ansatz solution.  Significant progress
has been made recently on this problem.  The breakthrough was the
realization that the Baxter $T$-$Q$ equation for this model should
have an inhomogeneous term \cite{Cao:2013qxa} (see also
\cite{Cao:2013nza}).  A simplified version of this solution was
subsequently shown to produce all the eigenvalues
\cite{Nepomechie:2013ila}.  A beautiful expression for the
corresponding eigenvectors was then proposed in
\cite{Belliard:2013aaa}.  Another simple solution was found and shown
to be complete in \cite{Jiang:2013rea}.

Despite these successes, an important question has remained
unanswered: is this solution practical for performing explicit
computations in the thermodynamic limit?  Due to the inhomogeneous
term in the $T$-$Q$ equation, the corresponding Bethe equations have a
non-conventional form; therefore, it appears that conventional Bethe
ansatz techniques for analyzing the thermodynamic limit (counting
function, root density, etc.)  cannot be used.

As a modest step towards addressing this question, we consider here
the problem of computing the so-called boundary (or surface) energy of
this model.  For simplicity, we focus on the limit that the
coefficient ($\xi$) of the non-diagonal boundary term goes to zero,
and compute the leading correction (of order $\xi^{2}$) to the
boundary energy.  In this limit, the $N$ Bethe roots for the ground
state split evenly into two sets: ``small'' roots that satisfy the
diagonal Bethe equations, and ``large'' roots that satisfy a
generalization of the Richardson-Gaudin equations.  The contribution
to the leading correction of the boundary energy from each of these
sets of roots can be evaluated exactly in the limit $N \rightarrow
\infty$.

The outline of this paper is as follows. In Section \ref{sec:model} 
we briefly describe the model and recall its Bethe ansatz solution.
In Section \ref{sec:boundaryenergy} we present the computation of the 
boundary energy. Our conclusions are presented in Section 
\ref{sec:conclusion}.

\section{The model and its Bethe ansatz solution}\label{sec:model}

We consider the antiferromagnetic open spin-1/2 isotropic quantum spin chain
with non-diagonal boundary terms.  Following \cite{Cao:2013qxa}, we
take as our Hamiltonian
\be
H =  \sum_{n=1}^{N-1}  
\vec \sigma_{n} \cdot \vec \sigma_{n+1} 
+ \frac{1}{q}\left(\sigma^{z}_{1} +\xi \sigma^{x}_{1}\right)
+ \frac{1}{p}\sigma^{z}_{N} \,,
\label{Hamiltonian}
\ee 
where $p, q, \xi$ are arbitrary real boundary parameters.  We consider the
solution based on the following linear $T$-$Q$ equation
\cite{Nepomechie:2013ila}:
\be
\Lambda(\lambda)\, Q(\lambda) 
= \bar{a}(\lambda)\, Q(\lambda-1) + \bar{d}(\lambda)\, Q(\lambda+1)
 + 2(1-\sqrt{1+\xi^{2}})\left( \lambda (\lambda+1)\right)^{2N+1} \,, 
\label{TQ}
\ee
where $\Lambda(\lambda)$ is an eigenvalue of the model's transfer matrix
\cite{Sklyanin:1988yz,deVega:1993xi,Cao:2013qxa}.  Moreover,
\be
\bar{a}(\lambda) = \frac{2\lambda+2}{2\lambda+1}(\lambda+ p)(\sqrt{1+\xi^{2}} \lambda +
q)(\lambda+1)^{2N} \,, \qquad \bar{d}(\lambda) = \bar{a}(-\lambda-1) \,,
\label{ad}
\ee  
and
\be
Q(\lambda) = 
\prod_{j=1}^{N}(\lambda-\lambda_{j})(\lambda+\lambda_{j}+1)  \,.
\label{Q}
\ee
The zeros $\lambda_{1}, \ldots, \lambda_{N}$ of $Q(\lambda)$ satisfy the 
Bethe equations that follow directly from (\ref{TQ}):
\be
\lefteqn{e_{1}(u_{j})^{2N}\, e_{2p-1}(u_{j})\, e_{2\tilde{q}-1}(u_{j})
-\prod_{\scriptstyle{k \ne j}\atop \scriptstyle{k=1}}^N 
e_{2}(u_{j}-u_{k})\, e_{2}(u_{j}+u_{k})} \label{BAE}\\ 
&&=i\left(1-\frac{1}{\sqrt{1+\xi^{2}}}\right)\frac{u_{j}(u_{j}+\frac{i}{2})^{2N}}{(u_{j}-i(p-\frac{1}{2}))
(u_{j}-i(\tilde{q}-\frac{1}{2}))\prod_{\scriptstyle{k \ne j}\atop 
\scriptstyle{k=1}}^N (u_{j}-u_{k}-i)(u_{j}+u_{k}-i)} \,, \non \\
&& 
\qquad\qquad\qquad\qquad\qquad\qquad\qquad\qquad\qquad\qquad\qquad\qquad\qquad\qquad\quad
j = 1, 2, \ldots, N \,,  \non 
\ee
where
\be
u_{j} = i\left(\lambda_{j}+{\textstyle\frac{1}{2}}\right)\,, \qquad \tilde{q} = \frac{q}{\sqrt{1+\xi^{2}}}\,, \qquad
e_{n}(u) = \frac{u + \frac{in}{2}}{u - \frac{in}{2}}\,.
\ee
The eigenvalues of the Hamiltonian (\ref{Hamiltonian}) are given by \cite{Nepomechie:2013ila}
\be
E = 
-2\sum_{j=1}^{N}\frac{1}{u_{j}^{2}+\frac{1}{4}} + 
N-1+\frac{1}{p}+\frac{1}{\tilde{q}}\,.
\label{energy}
\ee 

We observe that the energy is invariant under $\xi \rightarrow -\xi$,
since the $T$-$Q$ equation and Bethe equations have this invariance.
Moreover, we can restrict to one sign of $q$ (say, $q>0$), since the
signs of all the boundary terms in the Hamiltonian (\ref{Hamiltonian})
can be changed by a global $SU(2)$ transformation (namely, rotation by
$\pi$ about the $y$ axis, which leaves $\sigma^{y}$ invariant, but
changes $\sigma^{x,z} \rightarrow -\sigma^{x,z}$).  For definiteness,
we shall further restrict to even values of $N$, and $p<0$.

\section{Boundary energy}\label{sec:boundaryenergy}

For simplicity, we henceforth restrict our attention to the ground 
state. As is well known, for the corresponding closed chain Hamiltonian 
with periodic boundary conditions
\be 
H^{\rm periodic} =  \sum_{n=1}^{N}  
\vec \sigma_{n} \cdot \vec \sigma_{n+1} \,, \qquad \vec \sigma_{N+1} 
\equiv \vec \sigma_{1} \,,
\label{Hamiltonianperiodic}
\ee 
the ground-state energy $E_{0}^{\rm periodic}(N)$ for large $N$ is given by
\be
E_{0}^{\rm periodic}(N) = N e_{\infty} + O\left({\textstyle\frac{1}{N}}\right)\,,
\ee
where $e_{\infty}= 1-4 \ln 2$. In contrast, for the open chain 
Hamiltonian (\ref{Hamiltonian}), the 
ground-state energy $E_{0}(N;p,q,\xi)$ for large $N$ is given by (see, e.g. 
\cite{Gaudin:1971zza,Alcaraz:1987uk})
\be
E_{0}(N;p,q,\xi) = N e_{\infty} + E_{b}(p,q,\xi) + O\left({\textstyle\frac{1}{N}}\right)\,,
\ee
where $E_{b}(p,q,\xi)$ is the boundary (or surface) energy. Equivalently, we 
see that the boundary energy is given by
\be
E_{b}(p,q,\xi) = \lim_{N\rightarrow\infty} \left[ E_{0}(N;p,q,\xi) - E_{0}^{\rm 
periodic}(N) \right]
\,.
\ee 

The boundary energy is a function of the boundary parameters, and is
arguably the simplest such quantity to compute in the thermodynamic
limit.  For $\xi=0$, the boundary terms in the Hamiltonian
(\ref{Hamiltonian}) become diagonal, and the exact boundary energy is
known \cite{Hamer:1987ei,Batchelor:1989ja,Grisaru:1994ru,Kapustin:1995aw},
\be
E_{b}(p,q,\xi=0) &=&\frac{1}{p}+\frac{1}{q}-1+\pi
-2\int_{0}^{\infty}dx\, \frac{e^{-(2q-1)x}-e^{(2p-1)x}+e^{-x}}{\cosh 
x} \label{Ebxi0} \\
&=&\frac{1}{p}+\frac{1}{q}-1+\pi-\ln 4 
+\psi\big({\textstyle\frac{q}{2}}\big)-\psi\big({\textstyle\frac{1+q}{2}}\big)
+\psi\big({\textstyle\frac{2-p}{2}}\big)-\psi\big({\textstyle\frac{1-p}{2}}\big) 
\,, \non 
\ee
where $\psi(x)$ is the digamma function, and we have assumed that 
$q>0\,, p<0$. 

Unfortunately, the corresponding result for general values of $\xi$ is
still out of reach.  We therefore consider the series expansion of the
boundary energy about $\xi=0$,
\be
E_{b}(p,q,\xi) = E_{b}(p,q,\xi=0) + E_{b}^{(1)}(p,q)\, \xi^{2} + 
O\left(\xi^{4}\right)
\,,
\label{Ebexpansion}
\ee
which contains only even powers of $\xi$ since the energy is invariant
under $\xi \rightarrow -\xi$.  We focus here on computing only the
leading correction $E_{b}^{(1)}(p,q)$.

\begin{figure}
\centering
\subfloat[]{\includegraphics[width=5.5cm]{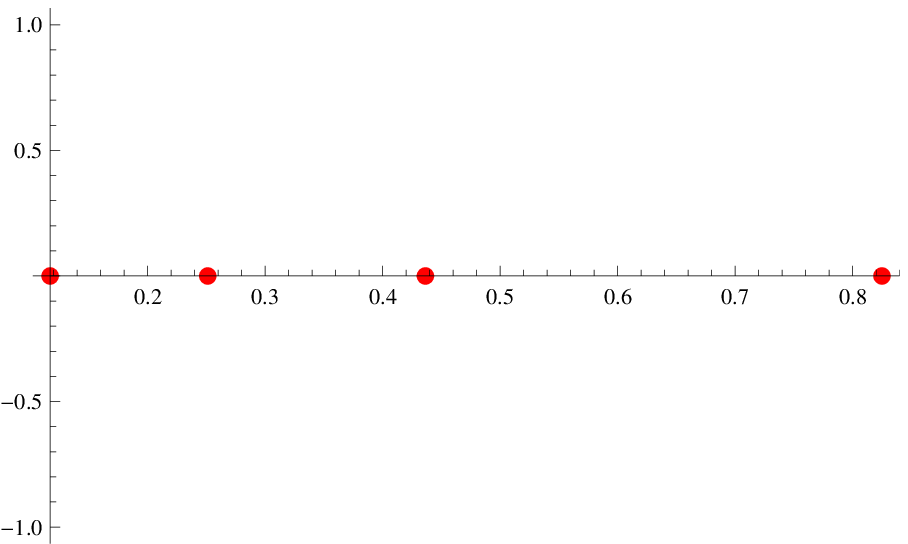}}
\hspace{2cm}
\subfloat[]{\includegraphics[width=5.5cm]{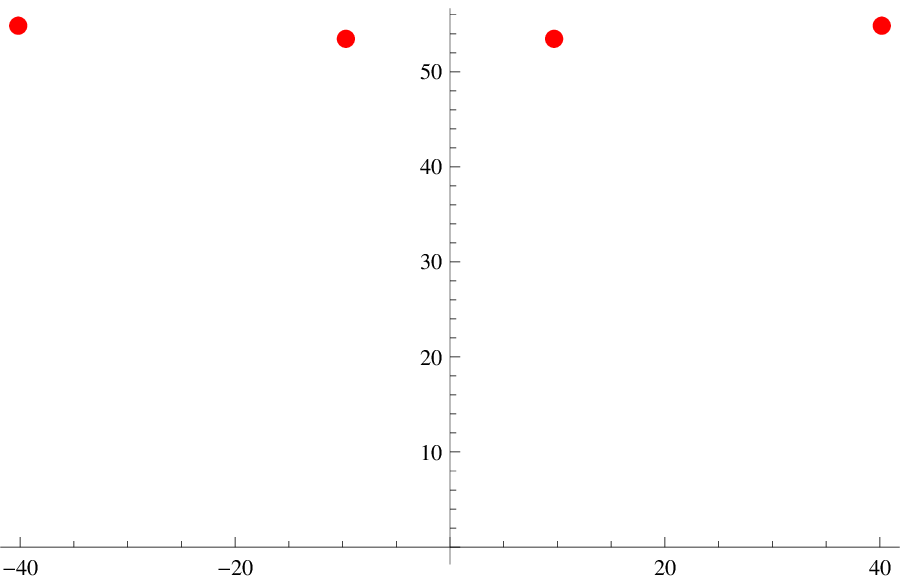}}
\caption{The exact small (a) and large (b) Bethe roots for the ground 
state with $p=-8\,, q=4\,, \xi=\frac{1}{8}\,, N=8$.} 
\label{fig:roots}
\end{figure}

From numerical studies for small values of $N$ (using the methods in
\cite{Nepomechie:2013ila}), we find that the $N$ Bethe roots $\{
u_{1}\,, \ldots \,, u_{N}\}$ describing the ground state split evenly
into two sets as $\xi \rightarrow 0$: ``small'' roots $\{ v_{1}\,,
\ldots \,, v_{N/2}\}$ that remain finite, and ``large'' roots $\{
w_{1}\,, \ldots \,, w_{N/2}\}$ that grow as $1/\xi$.  An example is 
shown in Fig. \ref{fig:roots}. We now proceed to consider
separately the contributions to $E_{b}^{(1)}(p,q)$ from these two sets
of roots.  

\subsection{Small roots}\label{subsec:small}

\begin{figure}
\centering
\subfloat[]{\includegraphics[width=5.5cm]{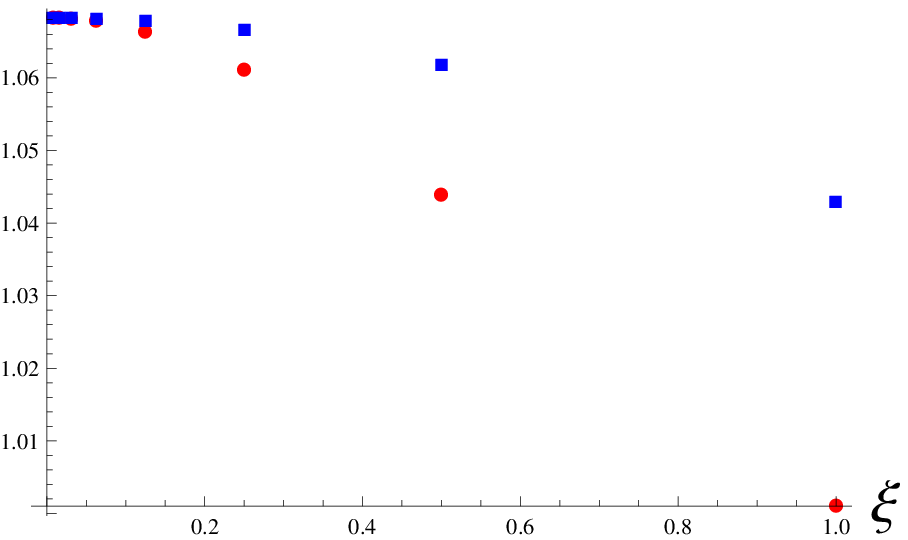}}
\subfloat[]{\includegraphics[width=5.5cm]{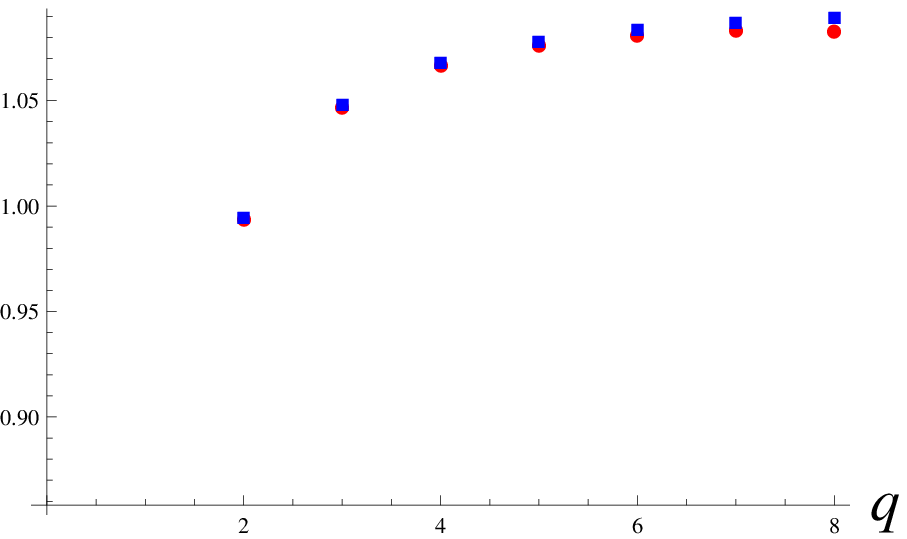}}
\subfloat[]{\includegraphics[width=5.5cm]{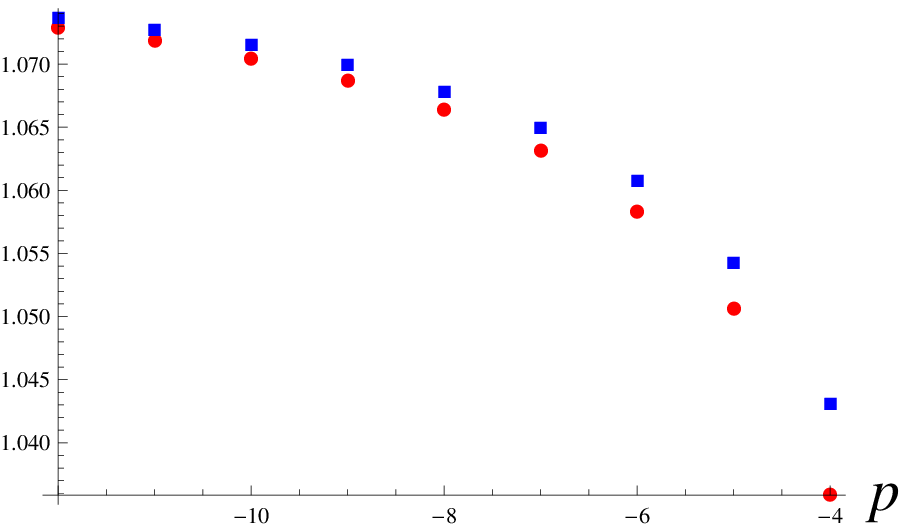}}
\caption{The boundary energy from the exact small roots ($E_{0}^{\rm 
small}(N)-E_{0}^{\rm periodic}(N)$)  is plotted with red circles; 
the boundary energy from the roots obtained using the diagonal Bethe 
equations (\ref{BAEsmall}) ($E_{0}^{\rm 
diag}(N)-E_{0}^{\rm periodic}(N)$) is plotted with blue squares. In 
(a), $p=-8\,, q=4$ and $\xi$ is varied; in (b), $p=-8\,, 
\xi=\frac{1}{8}$ and $q$ is varied; in (c), $q=4\,, 
\xi=\frac{1}{8}$ and $p$ is varied. In all three figures, $N=8$.} 
\label{fig:small_roots}
\end{figure}

For large values of $N$, we assume that the Bethe roots $\{ v_{1}\,, \ldots \,,
v_{N/2}\}$ that remain finite as $\xi \rightarrow 0$ decouple from the
large roots and approximately satisfy the diagonal reduction of the
exact Bethe equations (\ref{BAE}), namely,
\be 
e_{1}(v_{j})^{2N}\, e_{2p-1}(v_{j})\, e_{2\tilde{q}-1}(v_{j})
=\prod_{\scriptstyle{k \ne j}\atop \scriptstyle{k=1}}^{\frac{N}{2}}
e_{2}(v_{j}-v_{k})\, e_{2}(v_{j}+v_{k}) 
\,, \qquad j = 1\,, \ldots, \frac{N}{2} \,.  
\label{BAEsmall}
\ee 
These roots still depend on $\xi$ through $\tilde{q}$.  As a check on
this assumption, we have compared (for $N=8$, and for various values
of the boundary parameters) the boundary energy contributions from the exact
small roots, and from the Bethe roots obtained using the diagonal Bethe 
equations (\ref{BAEsmall}). We find that the agreement is very good
for small values of $\xi$, as shown in Fig. \ref{fig:small_roots}.

The contribution of these small roots to the boundary energy is given by
(\ref{Ebxi0}) with $q$ replaced by $\tilde{q}$.  Expanding this result
in powers of $\xi$, we recover the $\xi$-independent term
$E_{b}(p,q,\xi=0)$ in (\ref{Ebexpansion}), and we obtain from the term
of order $\xi^{2}$ the following contribution to $E_{b}^{(1)}(p,q)$
from the small roots:
\be 
E_{b}^{(1)\, {\rm small}}(p,q) =
\frac{1}{2q}-\frac{q}{4}\left[\psi'\left(\frac{q}{2}\right)-\psi'\left(\frac{q+1}{2}\right)\right]
\,.
\label{Eb1small}
\ee

\subsection{Large roots}\label{subsec:large}

For the Bethe roots $\{ w_{1}\,, \ldots \,, w_{N/2}\}$ that grow as
$1/\xi$ for $\xi \rightarrow 0$, we derive an approximate equation by
expanding the exact Bethe equations (\ref{BAE}) to first order in
$\xi$ using
\be
\frac{a+b}{a-b} = 1 + \frac{2b}{a} + O\Big(\Big(\frac{b}{a}\Big)^{2}\Big)\,, \qquad |a| \gg |b| \,.
\ee 
We obtain
\be
(p+q-1)\frac{1}{w_{j}}=\sum_{\scriptstyle{k \ne j}\atop \scriptstyle{k=1}}^{\frac{N}{2}}
\left(\frac{1}{w_{j}-w_{k}}+\frac{1}{w_{j}+w_{k}}\right) 
+{\textstyle\frac{1}{4}}\xi^{2} w_{j}\prod_{\scriptstyle{k \ne j}\atop \scriptstyle{k=1}}^{\frac{N}{2}}
\frac{1}{1-\left(\frac{w_{k}}{w_{j}}\right)^{2}} 
\,, \quad j = 1\,, \ldots, \frac{N}{2} \,. 
\label{BAElarge}
\ee 
These equations have some resemblance to those appearing in the
Richardson-Gaudin models \cite{Richardson:1964,Gaudin:1976sv}. 
However, the final term, which is due to the inhomogeneous term in 
the $T$-$Q$ equation (\ref{TQ}), is completely new.

It is hopeless to try to solve this equation directly, especially for
large values of $N$.  We proceed by instead recasting it in the form
of a $T$-$Q$-type equation, which however will be a differential
(rather than finite-difference) equation.  (Such a strategy has been
used for related problems in e.g.
\cite{Gaudin:1971zza,Shastry:2001,Lubcke:2004dg,Pan:2011,Marquette:2012}.)
To this end, we introduce the polynomial $\qq(w)$ of degree $N$ with
zeros $\pm w_{k}$,
\be
\qq(w) \equiv \prod_{k=1}^{\frac{N}{2}}(w-w_{k})(w+w_{k}) \,,
\ee 
which has the asymptotic behavior 
\be
\qq(w)  \sim w^{N} \quad  \mbox{  for   }\quad  w \rightarrow \infty \,.
\label{asymptotic}
\ee 
We observe the identities
\be
\frac{\qq''(w_{j})}{\qq'(w_{j})}=\frac{1}{w_{j}}
+2\sum_{k\ne 
j}\left(\frac{1}{w_{j}-w_{k}}+\frac{1}{w_{j}+w_{k}}\right) \,,
\ee 
and 
\be
\qq'(w_{j}) =  2 w_{j}^{N-1}\prod_{k\ne 
j}\left[1-\left(\frac{w_{k}}{w_{j}}\right)^{2}\right]
\,,
\ee 
where the prime denotes differentiation with respect to $w$.
It follows that (\ref{BAElarge}) is equivalent to
\be
w_{j} \qq''(w_{j}) -(2p+2q-1)\qq'(w_{j})  + \xi^{2} w_{j}^{N+1} = 0
\,,
\label{equivalent}
\ee 
The equation obtained by replacing $w_{j}$ with $-w_{j}$ in
(\ref{equivalent}) is consistent with (\ref{equivalent}), since
$\qq''(-w_{j}) = \qq''(w_{j})\,, \ \qq'(-w_{j}) = - \qq'(w_{j})$, and 
$N$ is even.  Therefore, the function 
\[ w \qq''(w) -(2p+2q-1)\qq'(w) + \xi^{2} w^{N+1} \] 
has all the zeros of $\qq(w)$, and is a polynomial of degree $N+1$.
It follows that
\be
w \qq''(w) -(2p+2q-1)\qq'(w)  + \xi^{2} w^{N+1} = t(w)\, \qq(w) \,,
\ee
where $t(w)$ is a polynomial of degree 1, i.e., $t(w)= t_{1} w + 
t_{0}$. The asymptotic behavior (\ref{asymptotic}) implies that 
$t_{1}=\xi^{2}\,, t_{0}=0$. We conclude that the zeros of $\qq(w)$ (and 
therefore the solutions of (\ref{BAElarge})) can 
be determined from the $T$-$Q$-type equation
\be
w \qq''(w) -(2p+2q-1)\qq'(w)  + \xi^{2} w^{N+1} - \xi^{2} w \qq(w) =0 \,.
\label{tq}
\ee
Remarkably, the unusual term in the Richardson-Gaudin-type equations 
(\ref{BAElarge}) (that originated from the inhomogeneous term in the 
$T$-$Q$ equation (\ref{TQ})) has been seamlessly accommodated.

\begin{figure}[h]
\begin{centering}
\includegraphics[width=7cm]{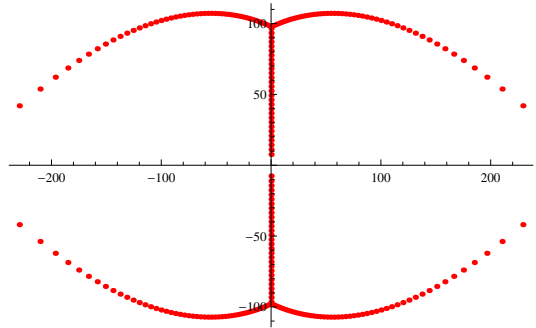}
\par\end{centering}

\caption{The zeros of $g(x)$ for $p=-8\,, q=4\,, N=256$.} 
\label{fig:gplot}
\end{figure}

Since the Bethe roots $\{ w_{j} \}$ grow as
$1/\xi$ for $\xi \rightarrow 0$, it is convenient to introduce
rescaled quantities
\be
x_{j} = w_{j} \xi \,, \qquad x = w \xi \,,
\ee
and the corresponding polynomial
\be
g(x) \equiv \prod_{k=1}^{\frac{N}{2}}(x-x_{k})(x+x_{k}) \,.
\ee
Evidently, $\qq(w) = \xi^{-N} g(x)$, and therefore (\ref{tq}) becomes
\be
x\frac{1}{g(x)}\frac{d^{2}g(x)}{dx^{2}}
-(2p+2q-1)\frac{1}{g(x)}\frac{dg(x)}{dx}+\frac{x^{N+1}}{g(x)} - x = 0 
\label{tqrescaled}
\,.
\ee 
Note that the $\xi$ dependence has disappeared.  This equation (or,
equivalently, Eq.(\ref{tq})) can be easily solved numerically for the
zeros of $g(x)$ even for large values of $N$, as shown in the example
of Fig.  \ref{fig:gplot}.

We observe that the term $\frac{x^{N+1}}{g(x)}$ in (\ref{tqrescaled})
goes to 0 for $x \sim 0$ and $N\rightarrow \infty$.  Indeed, $g(x)$
has no zeros near the origin (provided, as we henceforth assume, that
$p+q$ is not a positive integer), and therefore the denominator is
nonzero, while the numerator approaches zero rapidly for $x<1$ and $N
\rightarrow \infty$.  Hence, after dropping this term, the rescaled
$T$-$Q$-type equation (\ref{tqrescaled}) can be written as
\be
x\left( \frac{d G(x)}{dx} + G(x)^{2} -1\right) - (2p+2q-1) G(x) = 0 \,,
\qquad (x \sim 0 )\,,
\label{Geqn}
\ee 
where
\be
G(x) \equiv \frac{1}{g(x)}\frac{dg(x)}{dx} = 
\sum_{k=1}^{\frac{N}{2}}\left(\frac{1}{x-x_{k}}+\frac{1}{x+x_{k}}\right) \,.
\ee
The contribution of the large roots to the energy (\ref{energy}) can 
be expressed in terms of the derivative of $G(x)$ at $x=0$:
\be
E^{\rm large} = 
-2\sum_{j=1}^{\frac{N}{2}}\frac{1}{w_{j}^{2}+\frac{1}{4}} \approx 
-2\xi^{2}\sum_{j=1}^{\frac{N}{2}}\frac{1}{x_{j}^{2}}  = \xi^{2} 
\frac{d G(x)}{dx}\Big\vert_{x=0} \,.
\ee
The contribution of the large roots to $E_{b}^{(1)}(p,q)$ (\ref{Ebexpansion}) is therefore
\be
E_{b}^{(1)\, {\rm large}}(p,q) = \frac{d G(x)}{dx}\Big\vert_{x=0} \,.
\label{Eb1large1}
\ee

The first-order differential equation (\ref{Geqn}) can be solved in 
closed form
\be
G(x) = -i \frac{J_{p+q-1}(-i x)+ C Y_{p+q-1}(-i x) }{J_{p+q}(-i x) + C Y_{p+q}(-i x)}\,,
\label{Gsltn}
\ee
where $J_{n}(x)$ and $Y_{n}(x)$ are Bessel functions of the first and
second kind, respectively, and $C$ is an arbitrary constant.  The
requirement that $G(x)$ should be finite at $x=0$ uniquely determines 
$C$, which however depends on the value of $p+q$. For 
example, if $p+q\le 0$ and $p+q \ne -1/2$, then $C=0$. 

One way to evaluate (\ref{Eb1large1}) is to expand the Bessel
functions in (\ref{Gsltn}) about $x=0$ and obtain the $O(x)$ term.
Even easier is to substitute $G(x) = \alpha x + O(x^{2})$ into
(\ref{Geqn}) and solve for the constant $\alpha$.  We obtain
\be
E_{b}^{(1)\, {\rm large}}(p,q) = \frac{1}{2(1-p-q)} \,.
\label{Eb1large2}
\ee

\begin{figure}
\centering
\subfloat[]{\includegraphics[width=5.5cm]{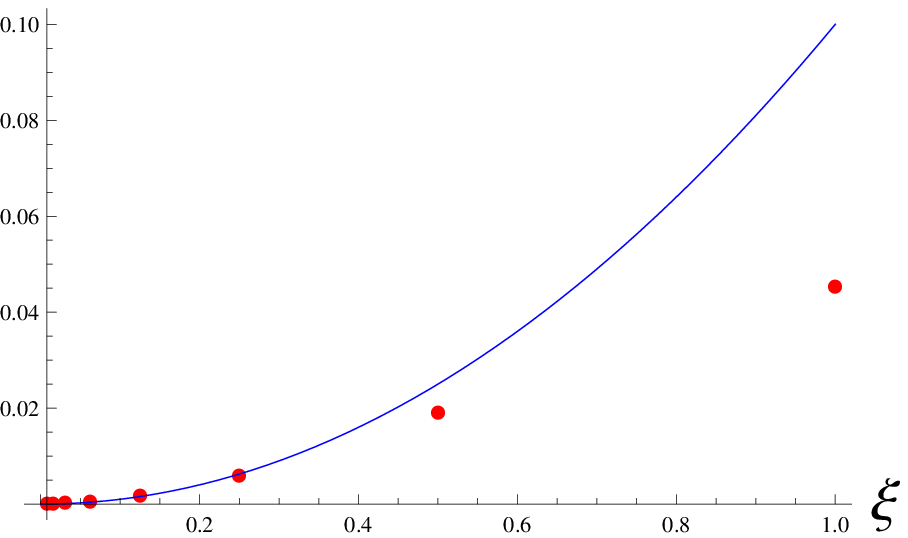}}
\subfloat[]{\includegraphics[width=5.5cm]{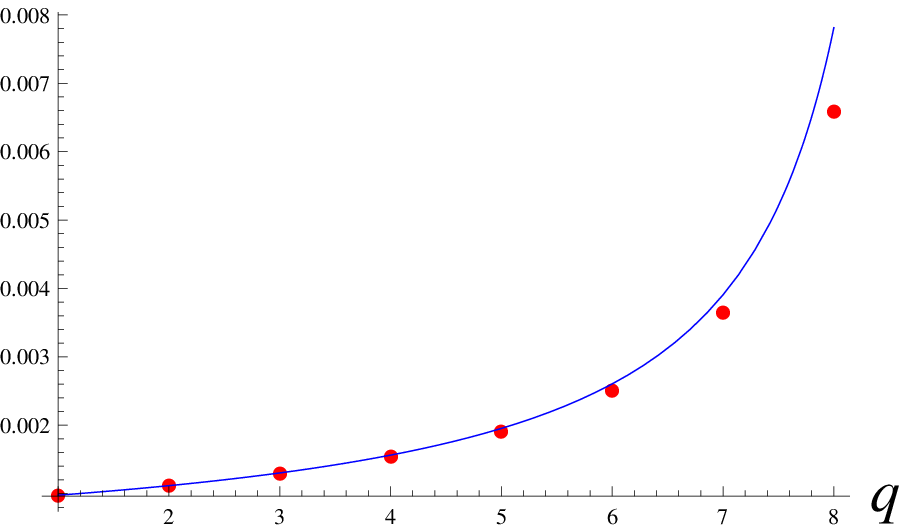}}
\subfloat[]{\includegraphics[width=5.5cm]{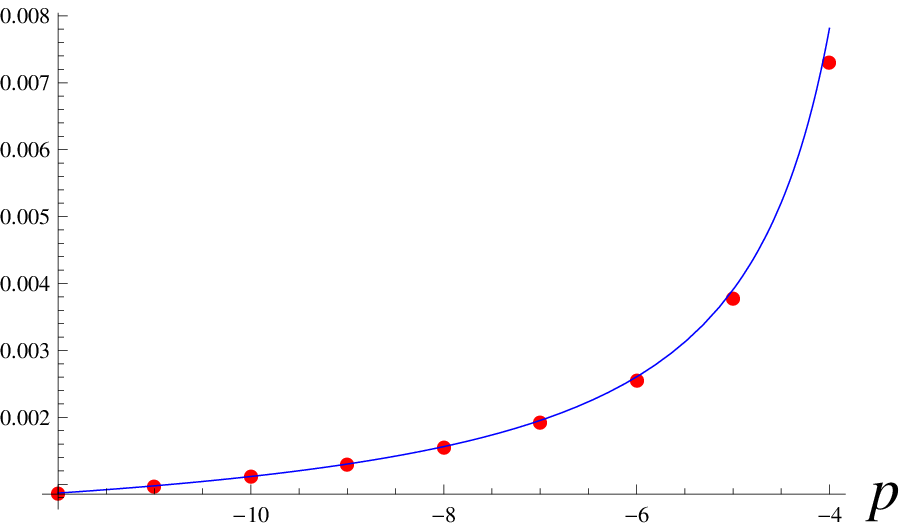}}
\caption{The energy from the exact large roots ($E_{0}^{\rm 
large}(N)$) for $N=8$ is plotted with red
circles; the $N\rightarrow \infty$ result $E_{b}^{(1)\, {\rm large}}(p,q)\,
\xi^{2}$, with $E_{b}^{(1)\, {\rm large}}(p,q)$ given by
(\ref{Eb1large2}), is the blue curve.  In (a), $p=-8\,,
q=4$ and $\xi$ is varied; in (b), $p=-8\,, \xi=\frac{1}{8}$ and $q$ is
varied; in (c), $q=4\,, \xi=\frac{1}{8}$ and $p$ is varied.}
\label{fig:large_roots}
\end{figure}

In deriving the result (\ref{Eb1large2}) for the contribution from the
large roots to the boundary energy, we have assumed that $N\rightarrow
\infty$.  Surprisingly, this result is accurate even for small values
of $N$ (provided that $\xi$ is small), as shown for $N=8$ in Fig.
\ref{fig:large_roots}.

\subsection{Final result}\label{subsec:total}

Adding the results from the small roots (\ref{Eb1small}) and the large 
roots (\ref{Eb1large2}), we obtain our final result for the 
leading correction to the boundary energy (defined in 
Eq. (\ref{Ebexpansion}))
\be
E_{b}^{(1)}(p,q) &=& E_{b}^{(1)\, {\rm small}}(p,q) + E_{b}^{(1)\, 
{\rm large}}(p,q) \non \\
&=& 
\frac{1}{2q}-\frac{q}{4}\left[\psi'\left(\frac{q}{2}\right)
-\psi'\left(\frac{q+1}{2}\right)\right] +\frac{1}{2(1-p-q)} \,.
\label{final}
\ee 

We have already noted in Figs.  \ref{fig:small_roots} and
\ref{fig:large_roots} some partial checks using numerical results for
$N=8$.  In principle, the final result (\ref{final}) could be checked
by comparing with numerical results for sufficiently large values of
$N$.  Indeed, boundary energies were estimated for the $\xi=0$
case in \cite{Alcaraz:1987uk} using extrapolation with values of $N$ up
to 256.  However, we have not (yet) managed to accurately solve the
exact Bethe equations (\ref{BAE}) numerically for the ground state
Bethe roots with such large values of $N$.

\section{Conclusion}\label{sec:conclusion}

We have argued that the recently-found Bethe ansatz solution
\cite{Cao:2013qxa,Nepomechie:2013ila} of the model (\ref{Hamiltonian})
can be used to perform a computation in the thermodynamic limit.
Indeed, at least for small values of $\xi$, the inhomogeneous term in
the $T$-$Q$ equation (\ref{TQ}), which leads to an unusual term in the
Richardson-Gaudin-type equations (\ref{BAElarge}) for the large roots,
does not impede the derivation of an analytical expression
(\ref{final}) for the boundary energy.  It would be interesting if one
could pass directly from the $T$-$Q$ equation (\ref{TQ}) to the
$T$-$Q$-type equation (\ref{tq}), without first going through the
equations (\ref{BAElarge}).

There are many interesting related problems: computing higher-order
corrections in $\xi$ and finite-size ($1/N$) corrections to the
ground-state energy, considering excited states, etc.  However, such
computations may require developing additional techniques.

\section*{Acknowledgments}
This work was supported in part by the National Science Foundation 
under Grant PHY-1212337, and by a Cooper fellowship.

\section{Erratum}\label{sec:intro}

We emphasized 
that our computation of the
boundary energy relies on several assumptions (in particular, the
decoupling of the ``small'' and ``large'' roots), and that the result
should therefore be checked numerically.  Although we have not managed
to solve the Bethe equations numerically for sufficiently large values
of $N$, we have succeeded to use the Density Matrix Renormalization
Group (DMRG) method, as implemented by the Algorithms and Libraries
for Physics Simulations (ALPS) \cite{ALPS}, to compute the
ground-state energy of the open chain up to 256 sites.  Sample results
are presented in Table \ref{table:numerical}. Following
\cite{Alcaraz:1987uk}, the large-$N$ extrapolation of the boundary
energy was performed using the van den Broeck-Schwartz algorithm
\cite{VBS, Hamer:1981, Henkel:1987ew} from the sequence $N=4, 6, 8,
\ldots, 60$. Note that $e_{\infty}= 1-4 \ln 2$.

\begin{table}[htb]
  \centering
  \begin{tabular}{|c|c|c|c|}\hline
    N &  $E_{0}(N)$ & $E_{0}(N) - N e_{\infty}$ \\
    \hline
     4 &     -6.50010714011 &   0.5902477488\\
     8 &    -13.5365249173  &   0.6441848606\\
     16&    -27.6843594238  &   0.6770601320\\
     32&    -56.0271138616  &   0.6957252501\\
     64&   -112.739844063   &   0.7058341603\\
     128&  -226.180208854   &   0.7111475927\\
     256&  -453.068824003   &   0.7138888904\\
     $\infty$ & &  0.7167\\
    \hline
   \end{tabular}
   \caption{Ground-state energy and boundary energy of the open chain 
   with boundary parameters $p=-8\,, q=4\,,  \xi=\frac{1}{8}$.}
   \label{table:numerical}
\end{table}

The extrapolated result for the boundary energy, $E_{b} =  0.7167$, is 
consistent with the contribution attributed to only the ``small'' roots,
\be
E_{b}(p,q,\xi=0) + E_{b}^{(1)\, {\rm small}}\, \xi^{2} \,,
\label{Eb}
\ee 
where $E_{b}(p,q,\xi=0)$ and $E_{b}^{(1)\, {\rm 
small}}$ are given by Eqs. (12) and (15) 
, respectively. 
Indeed, evaluating (\ref{Eb}) for our choice of boundary parameters gives 
0.716711, while adding the contribution from the ``large'' roots
\be
\frac{1}{2(1-p-q)} \xi^{2}
\label{Eblarge}
\ee
would yield a too-high value (0.718273) for the boundary energy. 

This result suggests that the boundary energy should be given entirely
by (\ref{Eb}) up to order $\xi^{2}$, and therefore by
\be
E_{b}=\frac{1}{p}+\frac{1}{\tilde{q}}-1+\pi
-2\int_{0}^{\infty}dx\, \frac{e^{-(2\tilde{q}-1)x}-e^{(2p-1)x}+e^{-x}}{\cosh 
x} \,, \qquad \tilde{q} = \frac{q}{\sqrt{1+\xi^{2}}} \,,
\label{Eball} 
\ee
for general values of $\xi$. Another argument for dropping the contribution
(\ref{Eblarge}) is that it remains constant in the limit $q
\rightarrow \infty$ and $p \rightarrow -\infty$ with $p+q$ constant,
which is inconsistent with the fact that the Hamiltonian becomes
independent of $\xi$ in this limit.  Moreover, (\ref{Eb}) and  
(\ref{Eball})
can be resolved into a sum of separate contributions from the two 
boundaries, as naively expected. The result (\ref{Eball}) has 
recently been derived by other means \cite{Li:2014}, see also 
\cite{Jiang:2013rea}.

The source of error in our computation is likely to be the assumption
of decoupling of the ``small'' and ``large'' roots. Unfortunately, it 
is not clear how to directly compute the contribution to the boundary 
energy due to the coupling of these two types of roots, which 
presumably should cancel (\ref{Eblarge}).

We are grateful to Junpeng Cao, Yupeng Wang and Wen-Li Yang for
valuable discussions, and to Shuai Cui for help with the DMRG
computations.

    
\providecommand{\href}[2]{#2}\begingroup\raggedright\endgroup

\end{document}